\documentclass{iopart}
\usepackage{graphicx}
\usepackage{amssymb}

\def\bbbr{{\Bbb R}}

\begin{document}

\title[The Kirchhoff Rod as a XY Spin Chain Model]{The Kirchhoff Rod as a XY Spin Chain Model}

\author{R Dandoloff$^\dag$ and G G Grahovski$^{\dag , \ddag}$}

\address{$^\dag$Laboratoire de Physique Th\'eorique et Mod\'elisation,
Universit\'e de Cergy-Pontoise, 2 avenue A. Chauvin, F-95302
Cergy-Pontoise Cedex,
FRANCE\\
$^\ddag$Laboratory of Solitons, Coherence and Geometry, Institute
for Nuclear Research and Nuclear Energy, Bulgarian Academy of
Sciences,72 Tsarigradsko chauss\'ee, 1784 Sofia, BULGARIA}
\ead{ggrahovs@ptm.u-cergy.fr $\qquad$ rossen@ptm.u-cergy.fr}

\begin{abstract}
A  XY Heisenberg spin chain model with two perpendicular spins par
site is mapped onto a Kirchhoff thin elastic rod. It is shown that
in the case of constant curvature the Euler--Lagrange equation
leads to the static sine-Gordon equation. The kink-antikink type
and periodical static solutions for these models are derived.

\end{abstract}

\submitto{\JPA}

\section{Introduction}\label{sec:1}

The study of an elastic rods \cite{love} is a subject to increased
interest \cite{ng1,gn1,mg1,db1,al,kdllt,Klapper} especially in
connection with the biomathematical models of proteins and of DNA
\cite{lya,fa}. The main feature of a thin rod is a space curve
(rod's axis) and the corresponding orthonormal frame with a
tangent vector ${\bf t}$ to the axial curve. The static energy of
the elastic rod is related to the bending and twisting energies.
It is tempting to map the elastic rod problem to a classical spin
chain \cite{ds1} (in the continuum limit, where the normalized
spin ${\bf S}$ is mapped onto the tangent vector $ {\bf t}$). We
will show however that the full mapping of the elastic rod onto a
spin-chain model requires a system of two orthogonal spins.

The spin Hamiltonian for a Heisenberg spin chain is given by the
following expression:
\begin{eqnarray}\label{eq:spin-Ham}
H=J_0 \sum_{i} {\bf S}_i(s)\cdot {\bf S}_{i+1}(s),\qquad {\bf
S}_{i}^2={\bf S}_{i+1}^2=1.
\end{eqnarray}
In the continuum limit this Hamiltonian goes over to
\begin{eqnarray}\label{eq:spin-Ham-cont}
H=J_0 \int_{-\infty}^{+\infty} \left( {d{\bf S}(x)\over
dx}\right)^2\, dx.
\end{eqnarray}
In the case of XY spin chain the spin is given by the rotation
angle $\theta (x)$: ${\bf S}(s)=(\cos \theta (x), \sin \theta
(x))$. The Hamiltonian now reads $H=J_0 \int_{-\infty}^{+\infty}
\left( {d\theta\over dx}\right)^2\, dx.$

This letter is organized as follows: In section 2 are discussed
the static properties of the Kirchhof equations for a thin elastic
rod. It is shown that if a curvature is present the twist angle
satisfies the static sine-Gordon equation. The mapping of the
Kirchhoff model onto a XY spin chain is done in Section 3. In
Section 4 the soliton-like solutions for the static sine-Gordon
equations are briefly discussed.

\section{Kirchhoff Model for Elastic Rods}\label{sec:Kirchhoff}

The model introduced by Kirchhoff (1859) describes the shape and
the dynamics of a thin elastic rod in equilibrium and is based on
the analogy with the dynamics of a heavy spinning top (the
Lagrange case). The shape is described by the {\it static
Kirchhoff model} while the time evolution -- by the {\it dynamical
Kirchhoff model}. Here we will concentrate  on the statics of thin
elastic rods (here and below we shall call them Kirchhoff rods).

We consider a static space curve ${\bf R}(s): \bbbr \rightarrow
\bbbr^3$ as a smooth function mapping the arc-length interval $I
\subset \bbbr$ into the physical space $\bbbr^3$. For every $s$ we
define the Frenet basis $({\bf t}(s), {\bf n}(s), {\bf b}(s))$ to
be the normal, binormal and the tangent vectors to the curve(s).
The tangent vector is a unit vector given by ${\bf t}=\left({
d{\bf R}\over ds}\right) $ and the curvature $\kappa (s) $ of the
curve at the point $s$ is then given by:
\[
\kappa (s):= \left|{d{\bf t}\over ds}\right|.
\]
The triad $({\bf t}(s), {\bf n}(s), {\bf b}(s))$ evolves along $s$
according to the Frenet--Serret equations:
\begin{eqnarray}\label{eq:fren-ser}
{d{\bf t}\over ds}=\kappa {\bf n}(s)\qquad {d{\bf n}\over
ds}=-\kappa {\bf t}(s)+\tau {\bf b}(s)\qquad {d{\bf b}\over
ds}=-\tau {\bf n}(s),
\end{eqnarray}
where $\tau $  is the torsion of the curve ${\bf R}(s)$. If the
curvature $\kappa$ and the torsion $\tau $ are known for all $s$
then the Frenet--Serret triad can be obtained as unique solution
of (\ref{eq:fren-ser}). Next the space curve ${\bf R}(s)$ can be
reconstructed by integrating the tangent vector ${\bf t}(s)$.

A thin rod can be modelled by a space curve ${\bf R}(s)$ joining
the loci of the centroids of the cross sections together with the
local basis $({\bf d}_1(s),{\bf d}_2(s),{\bf d}_3(s))$ attached to
the rod material. This local basis can be expressed through the
Frenet--Seret triad as follows:
\[
\left( {\bf d}_3(s), {\bf d}_2(s), {\bf d}_1(s)\right)=\left( {\bf
t}(s), {\bf n}(s), {\bf b}(s)\right) \left( \begin{array}{ccc}
1 & 0 & 0\\
0 & \cos \phi &-\sin \phi \\
0 & \sin \phi & \cos \phi \\
\end{array}
\right),
\]
where $\phi $ is the twist angle of the rod. The components of the
derivatives of the local basis $( {\bf d}_3(s), {\bf d}_2(s), {\bf
d}_1(s))$ with respect to $s$ can be expressed by using the twist
vector ${\bf k}(s)=\kappa_1{\bf d}_1+\kappa_2{\bf d}_2 +
\kappa_3{\bf d}_3$ as follows:
\[
{d {\bf d}_i\over ds} = {\bf k}(s) \times {\bf d}_i(s), \qquad
i=1,2,3.
\]
The static Kirchhoff equations describe the shape of the rod under
the effects of internal elastic stresses and boundary constraints,
in the absence of external force fields. Let ${\bf F}(s)$ is the
tension and ${\bf M}(s)$ is the torque of the rod. In the
approximation of a linear theory (the Hook's law applies) the
torque ${\bf M}$ is related to the twist vector ${\bf k}$ by ${\bf
M}(s)={\cal S}\cdot{\bf k}(s)$, where ${\cal S}=\mbox{diag} \,
(1,a,b)$. The constant $a$ measures the asymmetry of the cross
section and $b$ is the scaled torsional stiffness. In particular
for symmetric ($a=1$) hyperelastic ($b=1$) rods  we have ${\bf
M}(s)={\bf k}(s)$.  In the generic case the torque is
\begin{eqnarray}\label{eq:torque}
{\bf M}(s)= \kappa_1(s){\bf d}_1(s) + a\kappa_2(s){\bf d}_2(s) +
b\kappa_3(s){\bf d}_3(s)
\end{eqnarray}
and the elastic energy of the Kirchhoff rod is given by:
\begin{eqnarray}\label{eq:Helast}
H= {1\over 2} \int_{s_1}^{s_2} {\bf M}(s)\cdot {\bf k}(s)\, ds=
{1\over 2} \int_{s_1}^{s_2} (\kappa_1^2(s) + a\kappa_2^2(s) +
b\kappa_3^2(s))\, ds
\end{eqnarray}
The conservation of the linear and angular momenta is provided by
the static Kirchhoff equations:
\begin{eqnarray}\label{eq:Kirchhoff-f}
{d{\bf F}\over ds}=0, \\
\label{eq:Kirchhoff-m} {d{\bf M}\over ds}+ {\bf d}_3(s)\times {\bf
F}(s)=0.
\end{eqnarray}
Here ${\bf F}(s)$ is the tension of the rod and the torque ${\bf
M}(s)$ is given by (\ref{eq:torque}) and the twist vector reads
\begin{eqnarray}\label{eq:kappa_0}
{\bf k}(s)= (\kappa (s) \sin \phi , \kappa (s) \cos \phi , \tau +
\phi_s ),
\end{eqnarray}
where the twist angle $\phi $ is a function of the arc length
parameter $s$: $\phi=\phi (s)$.  The expression for the tension
${\bf F}(s) $ in the local basis ${\bf d}_1, {\bf d}_2, {\bf
d}_3$:
\[
{\bf F}(s) =F_1(s){\bf d}_1(s)+ F_2(s){\bf d}_2(s)+F_3(s){\bf
d}_3(s)
\]
reduces the Kirchhoff equations (\ref{eq:Kirchhoff-f}) and
(\ref{eq:Kirchhoff-m}) to the following system of ODE's:
\begin{eqnarray}\label{eq:Krichhoff-helix-f1s}
F_{1,s}+ \kappa_2 F_3 -\kappa_3 F_2=0
\\\label{eq:Krichhoff-helix-f2s} F_{2,s}+ \kappa_3 F_1 -\kappa_1
F_3=0 \\\label{eq:Krichhoff-helix-f3s} F_{3,s}+ \kappa_1 F_2
-\kappa_2 F_1=0 \\\label{eq:Krichhoff-helix-f1}
F_1=-a\kappa_{2,s}+
(b-1)\kappa_1\kappa_3\\\label{eq:Krichhoff-helix-f2}
F_2=\kappa_{1,s}+
(b-a)\kappa_2\kappa_3\\\label{eq:Krichhoff-helix-f3}
b\kappa_{3,s}+(a-1)\kappa_1\kappa_2=0.
\end{eqnarray}
Using the parameterization of the twist vector (\ref{eq:kappa_0})
from (\ref{eq:Krichhoff-helix-f3})  for the case of constant
curvature $\kappa(s)=\kappa_0 $ we get the famous static (scalar)
sine--Gordon equation:
\begin{eqnarray}\label{eq:Krichhoff-helix-sg}
{d^2 u \over ds^2} + {(a-1) \over b} \kappa_0^2 \sin u(s)=0,
\qquad u(s)=2\phi(s).
\end{eqnarray}
This second order differential equation is a completely integrable
Hamiltonian system and allows so-called ``soliton''-like
solutions. It appears in a wide variety of physical problems for
e.g. charge-density-wave materials, splay waves in membranes,
magnetic flux in Josephson lines, torsion coupled pendula,
propagation of crystal dislocations, Bloch wall motion in magnetic
crystals, two-dimensional elementary particle models in the
quantum field theory, etc.

For a symmetric rod, i.e. $a=1$ the equation
(\ref{eq:Krichhoff-helix-sg}) simplifies to ${d^2u\over ds^2}=0$,
i.e. $\phi_s=\mbox{const}$. This is the usual case widely
discussed in the literature \cite{ll7}. The cross-section of the
symmetric rod ($a=1$) has a  continuous rotational  symmetry
around the central axis. Therefore the elastic energy density $h$
does not depend on $\phi$ and from the variational principle it
follows that $h$ could be only a function of the derivatives
$\phi_s$. Therefore the only solution for a constant twist is
$\phi_s=\mbox{const}$.

There have been even attempts to generalize this result to the
asymmetric case ($a\neq 1$) \cite{mg1}, i.e. to show that
$\phi_s=\mbox{const}$ holds true for any Kirchhoff rod.

The asymmetric case has been overlooked for a long time.  For a
constant curvature and torsion along the centre-line it represents
another integrable case of the Kirchhoff equations for a thin
elastic rod. This opens new possibilities for a more adequate
modelling of bio-polymers and gives the phenomenological bases for
the widely used DNA models. Here there is no more continuous
rotational symmetry of the cross-section around the centre axis
and obviously $h$ depends on $\phi $ as well. So the constant
twist is no more a solution.

The solution of (\ref{eq:Krichhoff-helix-sg}) is compatible with
the full system of Kirchhoff  equations
(\ref{eq:Krichhoff-helix-f1s})--(\ref{eq:Krichhoff-helix-f3}) for
constant curvature and torsion.

\section{The Spin Chain Model}\label{sec:3}

Let us consider  the following spin chain model with two
perpendicular spins per site: the spin vectors ${\bf S}_1$ and
${\bf S}_2$ have different lengths and are given by
\begin{eqnarray}\label{eq:spin-vectors}
\fl {\bf S}_1 (s) &=& {\bf n}(s) \cos \phi + {\bf b}(s) \sin \phi, \qquad  {\bf S}_1^2=1, \qquad \phi=\phi(s); \\
\fl {\bf S}_2 (s) &=& -\sqrt{{b-a+1\over a+b-1}}{\bf n}(s) \sin
\phi + \sqrt{{b-a+1\over a+b-1}}{\bf b}(s) \cos \phi, \qquad  {\bf
S}_2^2={b-a+1\over a+b-1}. \nonumber
\end{eqnarray}
Here also both spin vectors are orthogonal:${\bf S}_1 (s)\cdot
{\bf S}_2 (s)=0$ for every $s$. This is an integrable system with
Hamiltonian given by
\[
H= J_0\sum_{i}\left( {\bf S}_{1,i} (s)\cdot {\bf S}_{1,i+1}(s) +
{\bf S}_{2,i}(s)\cdot {\bf S}_{2,i+1} (s)\right),
\]
which in the continuum limit leads to:
\begin{eqnarray}\label{eq:helix-hamilt}
\fl H= J_0\int_{s_1}^{s_2}\left[ \left({\partial {\bf S}_1  \over
\partial s} \right)^2 + \left({\partial {\bf S}_2 \over \partial
s} \right)^2\right]  \,ds= J_1\int_{s_1}^{s_2}\left( b(\tau
+\phi_s)^2 + \kappa^2(a-1)\sin ^2 \phi\right) \,ds, \nonumber
\end{eqnarray}
where $J_1$ is the renormalized coupling constant, $s\in
[s_1,s_2]$ and the subscript $s$ means a derivative with respect
to $s$. This Hamiltonian coincides with that one in
(\ref{eq:Helast}) where ${\bf k}(s)$ has been replaced from
(\ref{eq:kappa_0}). Thus the asymmetric Kirchhoff rod is mapped
onto a two-spin chain. If the curvature is constant
$\kappa(s)=\kappa_0$, then the Euler--Lagrange equation gives the
(scalar)  static sine--Gordon equation:
\begin{eqnarray}\label{eq:helix-sg}
 {d^2 \phi \over d s^2} + \kappa_0^2 {1-a\over b} \sin \phi \, \cos \phi =0.
\end{eqnarray}
\begin{figure}[htb]
\centerline{\includegraphics[width=0.7\textwidth]{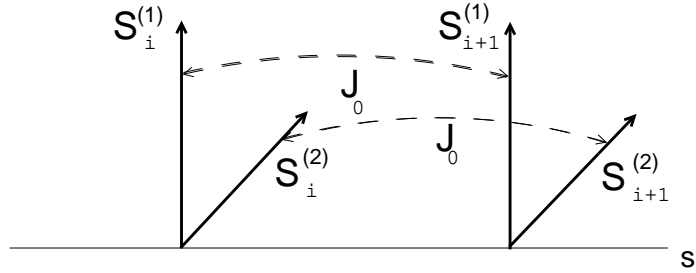}}
\caption{A two-spin XY chain system with a coupling constant $J_0$
that corresponds to the Kirchhoff rod model.}
 \label{fig:helix}
\end{figure}
When $a \to 1$ the Hamiltonian (\ref{eq:helix-hamilt}) simplifies
to $\int_{s_1}^{s_2} (\tau +\phi_s)^2\, ds $, so the
Euler-Lagrange equation leads to $\phi_{ss}=0$, or
$\phi_s=\mbox{const}$. Note that the case $a=1$ corresponds to a
symmetrical Kirchhoff rod. Such a rod-model is mapped onto a
symmetrical spin chain system $({\bf S}_1,{\bf S}_2)$ with ${\bf
S}_1^2={\bf S}_2^2=1$. Note that the mapping of the Kirchhoff
symmetric rod needs a two-spin XY chain rather than a simple
one-spin XY chain. From (\ref{eq:helix-sg}) one can easily get
that
\[
\left( {d \phi \over ds}\right)^2  + \kappa_0^2  {1-a\over b}
\sin^2 \phi (s) =0,
\]
which may be used for the calculation of the corresponding static
energy. For the asymmetric thin elastic rod  the model Hamiltonian
takes the form:
\begin{eqnarray}\label{eq:helix-model-hamilt}
H= \int_{s_1}^{s_2}\left( (\tau +\phi_s)^2 + \kappa_0^2{a-1\over
b}\sin ^2 \phi\right) \,ds
\end{eqnarray}
Here we shall discuss in brief  the static soliton solutions of
the model (\ref{eq:helix-model-hamilt}): The kink type solution of
the static sine-Gordon equation (\ref{eq:Krichhoff-helix-sg}),
(\ref{eq:helix-sg}) is given by
\begin{eqnarray}\label{eq:kink}
\phi(s)=2\,\mbox{arctan}\, \left[ \exp \left( {1\over \kappa_0}
\sqrt{{b\over 1-a}}s \right)\right] ,
\end{eqnarray}
and the corresponding static energy is
\begin{eqnarray}\label{eq:Energy-kink}
E_{\rm kink}= 4\kappa_0 \sqrt{{1-a\over b}}\, \mbox{tanh}\, \left(
\kappa_0 \sqrt{{1-a\over b}}l\right)
\end{eqnarray}
The periodic (soliton lattice) solution of
(\ref{eq:Krichhoff-helix-sg}), (\ref{eq:helix-sg}) is
\begin{eqnarray}\label{eq:periodic}
\phi(s)=2\, \mbox{arccos}\, \left[ \mbox{sn}\, \left(
{\kappa_0\over k}\sqrt{{1-a\over b}}s, k\right)\right]
\end{eqnarray}
with the periodicity $4{\kappa_0\over k}\sqrt{{1-a\over b}}K(k)$,
where $k$ is the modulus of the Jacobian elliptic function sn
(sine amplitude), and $K(k)$ is the complete elliptic integral of
the first kind. In the limit $k \to 1$ we have $K(k)\to \infty$
and the half-period tends to infinity as well, so we recover the
single kink soliton solution (\ref{eq:kink}).

The corresponding static energy per soliton of the soliton lattice
is given by:
\begin{eqnarray}\label{eq:E-periodic}
 E_{\rm soliton}={\kappa_0\over k}\sqrt{{1-a\over b}} \left(E(k)-{1\over 3}(k^{\prime})^2K(k) \right),
\end{eqnarray}
where $E(k)$ is the complete elliptic integral of second kind. In
the single soliton limit ($k \to 1$) the lattice energy per
soliton (\ref{eq:E-periodic}) reduces to eqn.
(\ref{eq:Energy-kink}).

\section{Conclusions}\label{sec:6}

We have shown that the single asymmetric elastic Kirchhoff rod
model can be mapped onto a 2-spin XY Heisenberg chain and the spin
vectors must have different lengths. In this case the
Euler-Lagrange equation for the spin chain Hamiltonian gives the
static sine-Gordon equation. For the case of symmetric  rods
($a=1$) both spins have the same length. The symmetric ($a=1$)
and the asymmetric ($a\neq 1$) Kirchhoff rods have very different
static properties. In general the family of thin elastic Kirchhoff
rods falls into two groups: i) the group of symmetric rods
($a=1$). Here the twist is constant along the rod and if the
torsion is constant as well, the Kirchhoff equations are
integrable and the curvature satisfies the non-linear
Schr\"odinger equation; ii) the group of asymmetric rods ($a \neq
1$). Here in general the twist and the curvature satisfy a coupled
differential equations (for a constant torsion). In the special
case where the curvature is constant the system of Kirchhoff
equations is integrable again and the twist satisfies the
sine-Gordon equation.

The dynamics of such models, which is of interest for realistic
biopolymers, should be investigated. Due to the Galilean
invariance of the sine-Gordon equation
(\ref{eq:Krichhoff-helix-sg})  a special class of dynamical
travelling wave type solutions can be obtained from the static
ones by Galilean boost.

The general assumption that all thin rods exhibit constant twist
should now be restricted to the class of symmetric thin rods only
and to all straight rods as well. The class of asymmetric thin
rods does not belong to this category. Here the twist is not
constant and ``interacts'' with the curvature. In the case of
constant curvature, the problem has exact solution. For
non-constant curvature the case is more complex and should be of
considerable interest e.g. for the problem of DNA supercoiling
 \cite{hh,blw,tt,yto,wto,sbh}.

\section*{Acknowledgments}
The work of GGG is supported by the Bulgarian National Scientific
Foundation Young Scientists Scholarship for the project
``Solitons, differential Geometry and Biophysical Models''. The
support by the National Science Foundation of Bulgaria, contract
No. F-1410 is also acknowledged.

\end{document}